\documentclass[prb,twocolumn,showkeys,showpacs,superscriptaddress,amsmath,amsfonts]{revtex4-1}
\usepackage[utf8x]{inputenc}
\usepackage{microtype}
\usepackage{lmodern}
\usepackage{graphicx}
\usepackage{hyperref}

\allowdisplaybreaks[1]

\newcommand{\FM}{\lvert\text{FM}\rangle}
\newcommand{\MF}{\langle\text{FM}\rvert}
\newcommand{\ket}[1]{\lvert#1\rangle}
\DeclareMathOperator{\Tr}{Tr}

\begin{document}

\title{Exact spectral function for hole-magnon coupling 
       in the ferromagnetic CuO$_3$-like chain}

\author{Krzysztof Bieniasz}
\email{krzysztof.bieniasz@uj.edu.pl}
\affiliation{Marian Smoluchowski Institute of Physics, Jagellonian University,
Reymonta 4, PL-30059 Krak\'ow, Poland}

\author{Andrzej M. Ole\'s}
\affiliation{Marian Smoluchowski Institute of Physics, Jagellonian University,
Reymonta 4, PL-30059 Krak\'ow, Poland}
\affiliation{Max-Planck-Institut f\"ur Festk\"orperforschung,
Heisenbergstrasse 1, D-70569 Stuttgart, Germany }

\date{\today}


\begin{abstract}
We present the exact spectral function for a single oxygen hole 
with spin opposite to ferromagnetic order within a one-dimensional 
CuO$_{3}$-like spin chain. We find that local Kondo-like exchange
interaction generates five different states in the strong coupling 
regime. It stabilizes a spin polaron which is a bound state of a 
moving charge dressed by magnon excitations, with essentially the 
same dispersion as predicted by mean field theory. We then examine 
in detail the evolution of the spectral function for increasing 
strength of the hole-magnon interaction. We also demonstrate that 
the $s$ and $p$ symmetry of orbital states in the conduction band 
are essentially equivalent to each other and find that the simplified 
models do not suffice to reproduce subtle aspects of hole-magnon 
coupling in the charge-transfer model.
\\
--- {\it Published in: Phys. Rev. B \textbf{88}, 115132 (2013).}
\end{abstract}

\maketitle

\section{Introduction}
\label{sec:intro}

The theoretical analysis of transition metal oxides, including cuprates,
manganites and iron pnictides, requires faithful description of strongly
correlated electrons which localize due to Coulomb interactions in
partly filled $3d$ orbitals.\cite{Ima98} These interactions lead to Mott 
insulators in undoped compounds, with spin and orbital degrees of 
freedom which interact with charge defects arising under doping
\cite{Zaa93} --- then the magnetic order and transport properties change 
due to subtle interplay between charge and magnetic/orbital degrees of 
freedom. Good examples are high temperature superconductivity in 
cuprates,\cite{Dag94,Nag06,Oga08} or colossal magnetoresistance in 
manganites.\cite{Dag01,*Dag05,Wei04,Tok06} In these systems, the interaction 
between charge carriers and localized spins is of crucial importance and 
drives the observed evolution of magnetic order and transport properties, 
captured in double exchange mechanism.\cite{deG60,vdB99,Tak02,Ole02,*Fei05} 
These changes may also depend on subtle quantum effects in systems with 
coupled spin-orbital-charge degrees of freedom.\cite{Ole12} 

A well known problem is the dynamics of one hole added to oxygen orbitals
which interacts with $S=1/2$ spins at Cu ions in CuO$_2$ planes of high 
temperature superconductors. The spins form an antiferromagnetic (AF) 
order due to the superexchange interaction.
A complete treatment of this problem involes a three-band model, 
\cite{Eme87,*Var87,*Ole87} with Cu $x^2-y^2$ orbitals occupied by one 
hole each and O $2p$ orbitals along the bonds. Instead, theoretical 
studies focus frequently on simplified treatments which do not include
all quantum effects related to charge carries interacting with spin 
excitations in phases with magnetic order. For example, following the 
idea of Zhang and Rice,\cite{Zha88} a simplified 
single-band model has been derived for CuO$_2$ planes from the 
charge-transfer model,\cite{Jef92,*Fei96,*Rai96} and next used to study
the evolution of magnetic order with increasing hole doping. However, 
such effective models do not accurately describe the electronic states in 
lightly doped materials. For instance, even low doping of less than 
5\% charge carriers is sufficient to change the magnetic order in
vanadates\cite{Fuj08,*Hor11} or in manganites.\cite{Sak10,*Ole11}

Electronic states change radically when electrons or holes propagate in
a background with magnetic order. The well known example is a single
hole which is classically confined in an antiferromagnet,\cite{Tru88}
but develops a quasiparticle propagating on the scale of superexchange
by its coupling to quantum spin fluctuations.\cite{Mar91} In contrast,
a conduction electron in the ferromagnetic (FM) background propagates
as a free particle, as known in FM semiconductors such as EuO or EuS.
\cite{Nol79} Here the electron spin oriented in the opposite way to the 
FM background scatters on magnon excitations which leads to rather 
complex many-body problem,\cite{Nol80,*Nol81} and to changes of the 
electronic structure with increasing temperature.\cite{Weg98} It was 
pointed out\cite{Nol96} that a repeated emission and reabsorption of a 
magnon by the conduction electron results in an effective attraction 
between magnon and electron. This gives rise to a polaron-like 
quasiparticle, the \emph{magnetic polaron}. 
Another excitation is due to a direct magnon emission or absorption by 
the electron, thereby flipping its own spin, leading to 
\emph{scattering states}. Modifications of electronic structure due to 
polarons were also discussed in manganites,\cite{Dag04} cobaltates,
\cite{Dag06} and vanadates.\cite{Ave13}

The purpose of this paper is to analyze the formation of polaron-like 
features and scattering states in a tight-binding model motivated
by the physical properties known from cuprates. Due to strong local
Coulomb repulsion $U$ at $x^2-y^2$ orbitals of Cu ions, the model
including holes in these orbitals and in the surrounding oxygen 
orbitals, called also a three-band model, reduces to a spin-fermion 
model.\cite{Zaa88,Pre88} The latter describes an oxygen hole coupled 
to the neighboring spins by a Kondo-like antiferromagnetic (AF) exchange
interaction. This local AF coupling frustrates the AF superexchange in 
CuO$_2$ planes and is responsible for a rapid decay of AF order under 
increasing doping. The main difficulty in treating the dynamics of a 
doped hole are the AF quantum fluctuations of the spin background, which 
have to be treated in an approximate way.\cite{Dag94,Nag06,Oga08}

Only very few many-body problems are exactly solvable. Exact solutions
are typically limited to one-dimensional (1D) models or to a very 
special choice of interaction parameters. However, an exact solution 
(i) provides always important physical insights into the nature of 
quantum states involved, 
(ii) could serve to test approximate treatments, and 
(iii) may be used to draw useful conclusions for experimental studies.
Recently, it was pointed out that a hole in a FM
system with a single magnon excitation provides valuable insights into
the spectral properties of a doped hole moving in a spin polarized
system.\cite{Mir12a,*Mir12b} Here we introduce a CuO$_3$-like spin-chain 
model, as studied for YBa$_2$Cu$_3$O$_7$ high temperature 
superconductors. Recently, excited states were investigated in AF CuO$_3$ 
chains\cite{Sch11} in Sr$_2$CuO$_3$ and an interesting interplay due to 
spin-orbital entanglement\cite{Ole12} was pointed out.\cite{Woh11} 
Here we analyze exactly the spectral properties in a FM chain. 
As we show below, they include the polaron-like and scattering 
states when the moving carrier interacts with magnons.

The paper is organized as follows. In Sec.~\ref{sec:model} we
introduce a 1D model for a CuO$_3$ spin chain.
The spectral function of a single charge added to the
oxygen orbital with the spin opposite to the FM order is obtained
exactly using the Green's function method in Sec.~\ref{sec:green}.
In Sec.~\ref{sec:app} we present an approximate perturbative solution 
for the same problem of a charge carrier coupled to the FM background 
in the strong coupling regime, while the mean field solution is given 
in Sec.~\ref{sec:mfa}. The
numerical results are presented in Sec.~\ref{sec:resu} and the exact
results are compared with the approximate ones.
Summary and conclusions are given in Sec.~\ref{sec:summa}, while
certain details of the derivation outlined in Sec.~\ref{sec:green}
are presented in the \hyperref[sec:appen]{Appendix}.

\section{The model}
\label{sec:model}

\begin{figure}[t!]
  \centering
  \includegraphics[width=\columnwidth]{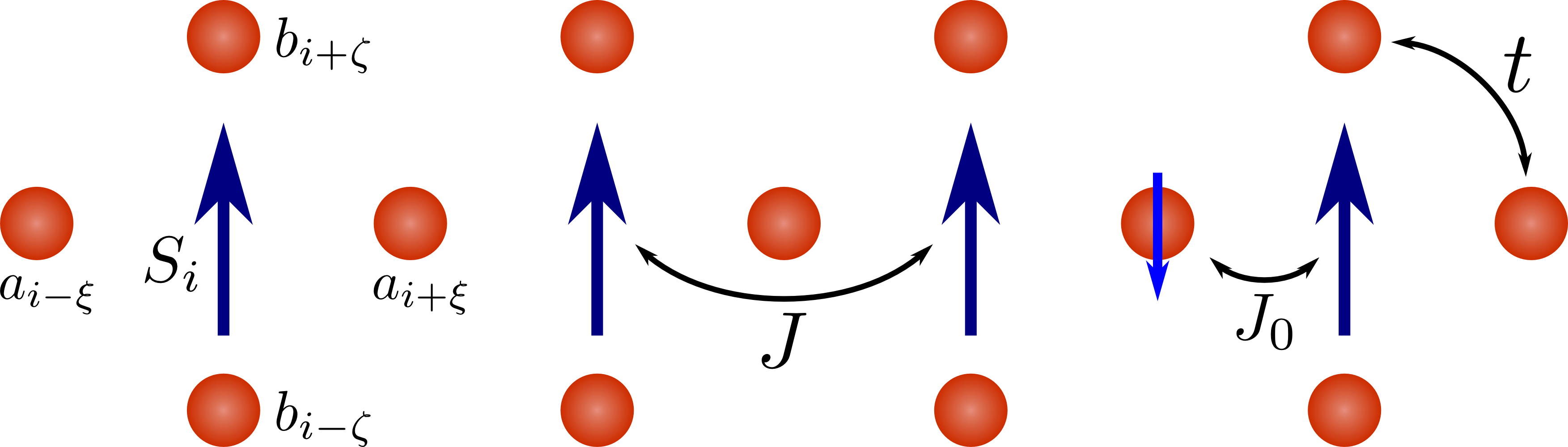}
\caption{
Graphic depiction of the CuO$_3$-like FM chain with localized spins
indicated by arrows and oxygen ions indicated by filled circles.
A hole added to an oxygen orbital (either $a_{i\pm\xi}$ or 
$b_{i\pm\zeta}$) interacts with the neighboring spin $S_{i}$ by the 
Kondo-like AF exchange $J_0$ while the localized spins interact by
the FM exchange $J$. The hole hops between 
neighboring oxygen orbitals by the hopping $t$. }
\label{fig:model}
\end{figure}

We consider a 1D model presented in Fig.~\ref{fig:model}, with the same
structure as a CuO$_3$ 1D chain in YBa$_2$Cu$_3$O$_7$, and assume that 
spins with a general value $S$ occupy the transition metal sites. In case 
of copper oxides, holes localize at Cu ions and $S=1/2$. Spins 
$\{\mathbf{S}_i\}$ are coupled here by FM Heisenberg exchange 
interactions as in the case of simpler 1D models considered before,
\cite{Mir12a,*Mir12b} while holes in oxygen orbitals represent charge 
degrees of freedom which couple to spins by a local AF exchange, similar 
to a hole added to a CuO$_2$ plane.\cite{Zaa88}
We label the oxygen orbitals as follows:
(i) $a_{i\pm\xi}$ is located in between the magnetic sites, where 
$\xi$ is a vector pointing from the Cu site towards the $a$ site
on its right, and
(ii) $b_{i\pm\zeta}$ is located above and below the magnetic sites, 
where $\zeta$ is a vector pointing from the Cu site towards the $b$ site 
above it.
Taking the charge-transfer model for a charged Cu$^{2+}$O$_3^{2-}$ 
chain as a reference (physical vacuum) state, these orbitals are filled 
with electrons and contain no hole.

The 1D model Hamiltonian,
\begin{equation}
  \label{eq:model}
  \mathcal{H} = \mathcal{T} + \mathcal{H}_{\mathrm{S}} + \mathcal{H}_{\mathrm{K}},
\end{equation}
includes the kinetic (hopping) part $\mathcal{T}$, the FM exchange
between localized spins, $\mathcal{H}_{\mathrm{S}}$, as well as Kondo-like
AF exchange interactions between a charge carrier (hole) in 
different orbitals and neighboring localized spins, $\mathcal{H}_{\mathrm{K}}$.
The hopping couples the $a$ and $b$ orbitals, 
see Fig.~\ref{fig:model}. Depending on the orbital symmetry, 
only one of the local combinations of $b$ orbitals contributes
to $\mathcal{T}$ and $\mathcal{H}_{\mathrm{S}}$, so it is convenient to 
introduce their symmetric ($+$, for $s$ orbitals) or antisymmetric
($-$, for $p$ orbitals) combinations, 
$b_{i}^{\pm}=(b_{i+\zeta} \pm b_{i-\zeta})/\sqrt{2}$.

The various terms in the Hamiltonian \eqref{eq:model} are:
\begin{subequations}
  \label{eq:hamilt1}
  \begin{align}
    \mathcal{T} &= -t\sum_{i\sigma}
    \left\{(a_{i+\xi,\sigma}^{\dag} \pm a_{i-\xi,\sigma}^{\dag})
      b_{i\sigma}^{\pm} + \text{H.c.}\right\},\\
    \mathcal{H}_{\mathrm{K}} &= J_{0}\sum_{i}
    \left(\mathbf{s}_{i+\xi}^{a}+\mathbf{s}_{i-\xi}^{a}
      +\mathbf{s}_{i}^{b}\right) \cdot \mathbf{S}_{i},\\
    \mathcal{H}_{\mathrm{S}} &= -J\sum_{i}(\mathbf{S}_{i}
    \cdot\mathbf{S}_{i+1}-S^{2}),
  \end{align}
\end{subequations}
where $\mathbf{S}_i$ is a spin operator for the magnetic ion at site 
$i$, $\mathbf{s}_m$ is a spin operator for the respective oxygen hole
in orbital $m=a,b$, and $S$ is the magnitude of a single localized spin 
on the magnetic sublattice. All the energy constants are positive 
($t>0$, $J_0>0$, $J>0$) and therefore $\mathcal{H}_{\mathrm{S}}$ 
provides FM coupling between the localized spins, while 
$\mathcal{H}_{\mathrm{K}}$ describes an AF Kondo-like coupling between
localized spins and conduction electrons.\cite{Zaa88}

We study below the dynamics of a single hole injected into either of 
the conduction bands, which arise after $\mathcal{T}$ is diagonalized 
--- one considers then two orbitals per unit cell and the Cu-Cu 
distance $a=1$. We will use the fermion 
representation for spin operators in the conduction band, 
$\mathbf{s}_{m}$. By transforming all the fermion operators to the 
reciprocal space by means of discrete Fourier transformation
one arrives at the following representation of the Hamiltonian,
\begin{subequations}
  \begin{align}
    \label{eq:hamilt2}
    \mathcal{T} &= \sum_{k\sigma}
    (\epsilon_{k}a_{k\sigma}^{\dag}b_{k\sigma}^{}+\text{H.c.}), \\
    \label{eq:hamilt2ex}
    \mathcal{H}_{\mathrm{K}} &= J_{0}\sum_{kq}
    [2\cos(q/2)\,\mathbf{s}_{kq}^{a}+\mathbf{s}_{kq}^{b}]\cdot\mathbf{S}_{q},
  \end{align}
\end{subequations}
where $\epsilon_{k}$ follows from the Fourier fransformation and is 
given by 
\begin{equation}
  \label{eq:dispersion}
  \epsilon_{k}=
  \begin{cases}
    -2t\cos(k/2) & \text{for $s$ symmetry}\\
    +2it\sin(k/2) & \text{for $p$ symmetry}
  \end{cases}.
\end{equation}
This leads to two bands $\varepsilon_k=\pm|\epsilon_k|$ for each value 
of $k\in[-\pi,\pi)$. The reciprocal-space spin operators are given by:
\begin{align}
  \label{eq:spin}
  \mathbf{S}_{q} &= \frac{1}{N}\sum_{i} e^{-iqR_{i}}\,\mathbf{S}_{i} ,\\
  \mathbf{s}_{kq}^{\mu} &=
  \begin{pmatrix}
    \tfrac{1}{2}(\mu_{k\uparrow}^{\dag}\mu_{k+q\uparrow}
-\mu_{k\downarrow}^{\dag}\mu_{k+q\downarrow})\\
    \mu_{k\uparrow}^{\dag}\mu_{k+q\downarrow}\\
    \mu_{k\downarrow}^{\dag}\mu_{k+q\uparrow}
  \end{pmatrix}=
  \begin{pmatrix}
    \text{$s^{z}_{kq}$}\\
    \text{$s^{+}_{kq}$}\\
    \text{$s^{-}_{kq}$}
  \end{pmatrix},
\end{align}
where $\mu$ is an index labeling the states $\{a,b\}$. It should be
emphasized that, strictly speaking, the operators $\mathbf{s}_{kq}^{\mu}$ 
are just a shorthand notation for the respective fermionic operators 
and should not be confused with regular spin operators. However, their 
effect in the spin subspace is similar.

As for $\mathcal{H}_{\mathrm{S}}$, its following eigenstates are easily
identified:
\begin{align}
  \label{eq:Hh0}
  \mathcal{H}_{\mathrm{S}} \FM &= 0 \FM,\\
  \label{eq:HhSq}
  \mathcal{H}_{\mathrm{S}} S_{q}^{-} \FM &= \Omega_{q} S_{q}^{-} \FM,
\end{align}
where $\FM$ is the physical vacuum state, and $S_{q}^{-}\FM$, defined
by Eq.~\eqref{eq:spin}, is a magnetic excited state with one magnon
(spin wave) created in the FM background and its energy dispersion
\begin{equation}
  \label{eq:mag}
  \Omega_{q}=4JS\sin^{2}(q/2). 
\end{equation}
Since the Hamiltonian under
consideration conserves the total spin, these magnon states are the
only attainable in the problem of a single hole with spin
$s=1/2$ coupled to the FM spin background.

\section{Spin polaron and scattering states}

\subsection{Exact solution by Green's functions}
\label{sec:green}

To obtain the hole spectral function we calculate first the Green's 
function, defined by the expectation value of the resolvent,
\begin{equation}
  \label{eq:resolv}
  \mathcal{G}(\omega) = [\omega-\mathcal{H}+i\eta]^{-1},
\end{equation}
for the $\downarrow$-spin states of an added hole. Therefore, the
Green's function has a $2 \times 2$ matrix structure:
\begin{equation}
  \label{eq:green}
  \mathbb{G}_{\mu\nu}(k,\omega) =
\MF\mu_{k\downarrow}\mathcal{G}(\omega)\nu_{k\downarrow}^{\dag}\FM,
\end{equation}
where $\mu,\nu$ are again indices going over the states $\{a,b\}$.
Following a method similar to the one described by Berciu and Sawatzky,
\cite{berciu09} we divide the Hamiltonian, 
$\mathcal{H}_{0}=\mathcal{T}+\mathcal{H}_{\mathrm{S}}$, into the free 
part corresponding to $\mathcal{G}_{0}$, and the term 
$\mathcal{V}=\mathcal{H}_{\rm K}$ which couples the two subsystems by
the AF interaction $\propto J_0$.

It is convenient to represent the Hamiltonian \eqref{eq:hamilt2} in
terms of the following matrices:
\begin{subequations}
  \label{eq:matrix}
  \begin{align}
    \mathbb{T}(k) &=
    \begin{pmatrix}
      0 & \epsilon_{k}\\
      \epsilon_{k}^{*} & 0
    \end{pmatrix},\\
    \label{eq:vmatrix}
    \mathbb{V}(q) &=
    \begin{pmatrix}
      \cos(q/2) & 0\\
      0 & \tfrac{1}{2}
    \end{pmatrix},
  \end{align}
\end{subequations}
while the form of $\mathbb{T}$ leads us to the matrix representation of
$\mathcal{G}_{0}(\omega)$:
\begin{equation}
  \label{eq:G0}
  \mathbb{G}_{0}(k,\omega) =
  \begin{pmatrix}
    \omega+i\eta & -\epsilon_{k}\\
    -\epsilon_{k}^{*} & \omega+i\eta
  \end{pmatrix}^{-1},
\end{equation}
and in the case of one magnon state \eqref{eq:HhSq}, the magnon energy 
$\Omega_{q}$ \eqref{eq:mag} is taken into account by substituting 
$\omega\to\omega-\Omega_{q}$. The inverse could also be calculated 
explicitly; however, it is not necessary for the present derivation.

We then proceed by utilizing the Dyson's equation,
\begin{equation}
  \label{eq:dyson}
  \mathcal{G}(\omega) = \mathcal{G}_{0}(\omega)
  + \mathcal{G}(\omega)\mathcal{V}\mathcal{G}_{0}(\omega),
\end{equation}
which, after separating $\mathbb{G}(k,\omega)$, leads to the following 
matrix equation,
\begin{equation}
  \label{eq:G1}
  \mathbb{G}(k,\omega) = [\mathbb{I} + J_{0}\mathbb{F}(k,\omega)]
  \mathbb{G}_{0}(k,\omega) \mathbb{Q}_{G}(k,\omega),
\end{equation}
where the various auxiliary matrices are given by:
\begin{align}
  \label{eq:fmatrix}
  \mathbb{F}(k,\omega) &= \sum_{q}\mathbb{\tilde{F}}(k,q,\omega)\mathbb{V}(q),\\
  \label{eq:anomalous}
  \mathbb{\tilde{F}}_{\mu\nu}(k,q,\omega) &=
  \MF \mu_{k\downarrow}\mathcal{G}(\omega)\nu_{k-q,\uparrow}^{\dag} S_{q}^{-} \FM.
\end{align}
Here $\mathbb{\tilde{F}}(k,q,\omega)$ is the anomalous Green's function,
calculated between different magnon states, resulting from the $S^{-}$
terms in $\mathcal{V}$, and
\begin{align}
  \label{eq:qmatrix}
  \mathbb{Q}_{G}(k,\omega) &= \big[\mathbb{I}
    +J_{0}S\mathbb{V}_{0} \mathbb{G}_{0}(k,\omega)\big]^{-1},\\
  \mathbb{V}_{0} &= \sum_{q} \mathbb{V}(q)\delta_{q0} =
  \begin{pmatrix}
    1 & 0 \\
    0 & \frac{1}{2}
  \end{pmatrix},
\end{align}
where $\mathbb{Q}_{G}(k,\omega)$ is a transformation of 
$\mathbb{G}_{0}(k,\omega)$, performing a constant shift by $J_{0}S$.
However, this cannot be written shortly as
$\mathbb{G}_{0}(k,\omega+J_{0}S)$, because of the matrix $\mathbb{V}_0$ 
present in $\mathbb{Q}_{G}(k,\omega)$, which causes a different shift 
of ${J_{0}S}/{2}$ in the $G_{0}^{bb}(k,\omega)$ sector.

The next step is to eliminate $\mathbb{F}(k,\omega)$ from Eq.~\eqref{eq:G1}.
In order to do this, one needs to express 
$\mathbb{\tilde{F}}(k,\omega)$ explicitly in terms of 
$\mathbb{G}(k,\omega)$ by applying the Dyson's equation~\eqref{eq:dyson} 
once again and next solving for $\mathbb{F}(k,\omega)$. After inserting 
it back into Eq.~\eqref{eq:G1} and solving for $\mathbb{G}(k,\omega)$, 
one arrives at the final result,
\begin{widetext}
  \begin{equation}
    \label{eq:G2}
    \mathbb{G}(k,\omega) = \mathbb{G}_{0}(k,\omega)\mathbb{Q}_{G}(k,\omega)
    \Big[\mathbb{I}-2J_{0}S \big( \mathbb{I} - \mathbb{M}^{-1}(k,\omega) \big)
    \mathbb{G}_{0}(k,\omega)\mathbb{Q}_{G}(k,\omega) \Big]^{-1},
  \end{equation}
\end{widetext}
where $\mathbb{M}$ is a complicated matrix expressed solely,
in terms of various sums of
$\mathbb{G}_{0}(k-q,\omega-\Omega_{q})$ over $q$. More details are
presented in the \hyperref[sec:appen]{Appendix}. We note that this
solution is almost identical to the one obtained by Berciu in Ref. 
\onlinecite{berciu09}, only here we arrive at a more general solution 
for the transformation of $\mathbb{G}_{0}(k,\omega)$.
Finally, having calculated the Green's function, one finds the
spectral function,
\begin{equation}
  \label{eq:spectral}
  \mathbb{A}(k,\omega) = -\frac{1}{\pi}\Im{[\mathbb{G}(k,\omega)]},
\end{equation}
which is closely related to the density of states
as well as to the photoemission spectra, and can be directly measured
in angle resolved photoemission spectroscopy experiments. The main 
physical problem is its structure and possible quasiparticle (QP) states.

The Green's function $\mathbb{G}(k,\omega)$, as calculated from
Eq.~\eqref{eq:G2}, is generally not diagonal. This is usually not a
problem, since both diagonal components of the spectral function are 
measured at once in experiment, which corresponds to the
trace of the corresponding matrix~\eqref{eq:spectral},
\begin{equation}
  \label{eq:trace}
  A(k,\omega) = \Tr \mathbb{A}(k,\omega),
\end{equation}
a quantity invariant under the change of basis. Thus, we also
present here the traced spectral function $A(k,\omega)$.
In order to get more physical insight into the exact solution, we will 
now derive the approximate solutions of the problem, in two opposite 
parameter regimes, strong and weak hole-magnon coupling. 

\subsection{Perturbative solution at strong coupling}
\label{sec:app}

First approach is the perturbation expansion, with the problem treated 
in the eigenbasis of $\mathcal{V}$. This solution is valid in the 
strong coupling limit $J_0\gg t$ and $J_0\gg J$, since we treat 
$\mathcal{T}$ and $\mathcal{H}_{\mathrm{S}}$ as small perturbations to 
$\mathcal{H}_{\mathrm{K}}$.

Given the conjectured states of the form $\mu_{k\downarrow}^{\dag}\FM$
and $\sum_{q} e^{iq\rho} \nu_{k-q,\uparrow}^{\dag}S_{q}^{-}\FM$ (where 
$\rho=0,\pm\xi$ depends on the specific orbital) a straightforward 
calculation shows that the eigenstates of $\mathcal{V}$ are:
\begin{widetext}
\begin{subequations}
  \label{eq:eigstate}
  \begin{align}
    \ket{\bar{a}}_{k} &= \sqrt{\frac{4S}{4S+1}}
\left[a_{k\downarrow}^{\dag} -\frac{1}{2S}\sum_{q}\cos\left(\frac{q}{2}\right)
a_{k-q,\uparrow}^{\dag} S_{q}^{-}\right] \FM,\\
    \ket{\bar{b}}_{k} &= \sqrt{\frac{2S}{2S+1}}
\left[b_{k\downarrow}^{\dag} - \frac{1}{2S} \sum_{q}
b_{k-q,\uparrow}^{\dag} S_{q}^{-}\right] \FM,\\
    \ket{m}_{k} &= \frac{1}{\sqrt{S}} \sum_{q}\sin\left(\frac{q}{2}\right)
a_{k-q,\uparrow}^{\dag} S_{q}^{-} \FM,\\
    \ket{b}_{k} &= \frac{1}{\sqrt{2S+1}} \left[b_{k\downarrow}^{\dag}
+ \sum_{q} b_{k-q,\uparrow}^{\dag} S_{q}^{-}\right] \FM,\\
    \ket{a}_{k} &= \frac{1}{\sqrt{4S+1}} \left[a_{k\downarrow}^{\dag} 
+2\sum_{q}\cos\left(\frac{q}{2}\right)a_{k-q,\uparrow}^{\dag}S_q^-\right]\FM.
  \end{align}
\end{subequations}

The first two states are bound polaronic states, and the last two ones 
are the respective excited states. The remaining state $\ket{m}_{k}$ is 
a state dominated by magnons which dress an $\uparrow$-spin hole that  
propagates over $a$ orbitals. These definitions allow us to infer 
something about the approximate nature of different bands calculated 
from the Green's function. Further, because $\mathcal{V}$ does not 
involve any three-site interaction terms but only self-renormalizing 
exchange interaction, there is no distinction between $s$ and $p$ 
orbitals, and therefore in both cases the states derived in perturbation 
theory~\eqref{eq:eigstate} are the same. 
Using them, one can calculate the perturbation corrections to their 
energy coming from $\mathcal{H}_{\mathrm{S}}$ and $\mathcal{T}$. Owing 
to the specific orbital symmetries in the latter, in order to get a
nontrivial contribution (i.e., dispersion) one needs to conduct the 
perturbation expansion at least up to the second order. The resulting
energies for the states \eqref{eq:eigstate} are, respectively:
  \begin{subequations}
    \label{eq:eigenergy}
    \begin{align}
      E_{\bar{a}} &= -J_{0}\frac{(2S+1)}{2} +J\frac{2S}{4S+1}
      -\frac{\epsilon_{k}^{2}}{J_{0}}\frac{1}{(2S+1)}
      \left[\frac{(4S+1)}{S}+\frac{2S}{(4S+1)(3S+1)}\right]\,,\\
      E_{\bar{b}} &= -J_{0}\frac{S+1}{2} +J\frac{2S}{2S+1}
      +\frac{\epsilon_{k}^{2}}{J_{0}}\frac{4S+1}{S(2S+1)}
      -\frac{\epsilon_{k-\pi}^{2}}{J_{0}}\frac{1}{3S(2S+1)}\,,\\
      E_{m} &= J_{0}\frac{2S-1}{2} +J2S
      +\frac{\epsilon_{k-\pi}^{2}}{J_{0}}\frac{1}{2S+1}
      \left[\frac{1}{3S}-\frac{2S}{1-S}\right]\,,\\
      E_{b} &= J_{0}\frac{S}{2} +J\frac{4S^{2}}{2S+1}
      +\frac{\epsilon_{k}^{2}}{J_{0}}\frac{2}{4S+1}
      \left[\frac{S}{(2S+1)(3S+1)}-\frac{2S+1}{S}\right]
      +\frac{\epsilon_{k-\pi}^{2}}{J_{0}}\frac{2S}{(2S+1)(1-S)}\,,\\
      E_{a} &= J_{0}S +J\frac{8S^{2}}{4S+1}
      +\frac{\epsilon_{k}^{2}}{J_{0}}\frac{2(2S+1)}{S(4S+1)}\,.
    \end{align}
  \end{subequations}
\end{widetext}

\subsection{Mean field approximation}
\label{sec:mfa}

Another approximate approach to the problem is the mean field (MF)
approximation for $\mathcal{V}$. In this case the principle is to
neglect the quantum spin fluctuations in $\mathcal{H}_{\mathrm{K}}$,
effectively setting $\mathbf{s}_m\cdot\mathbf{S}_i\approx s^z_mS_i^z$,
where $\mathbf{s}_m$ stands for a spin in itinerant orbital, $a$ or $b$, 
in the neighborhood of site $i$.
This assumption is valid provided the whole $\mathcal{H}_{\rm K}$ brings
only a minor contribution to the overall energy, therefore implying
$J_0\ll t$ and $J_0\ll J$. From Eq.~\eqref{eq:anomalous} it follows that 
neglecting spin fluctuations implies $\tilde{\mathbb{F}}(k,q,\omega)=0$, 
and thus~\eqref{eq:G1} reduces to the MF solution of the Green's
function,
\begin{equation}
  \label{eq:GMF}
  \mathbb{G}_{MF}(k,\omega) = \mathbb{G}_{0}(k,\omega) \mathbb{Q}_{G}(k,\omega).
\end{equation}
This equation depends only on $\mathbb{G}_{0}(k,\omega)$ and can
be solved analytically, yielding the mean field energy dispersion,
\begin{equation}
  \label{eq:MF}
  E_{MF}^{\pm}(k) = -\frac{3J_{0}S}{4}
  \pm\sqrt{\left(\frac{J_{0}S}{4}\right)^{2}+\epsilon_{k}^{2}}.
\end{equation}
This is also an exact solution of the model~\eqref{eq:model} with Ising
interactions in $\mathcal{H}_{\rm K}$, and the deviation from it,
reported in Sec.~\ref{sec:resu}, is due to quantum spin fluctautions. 

Furthermore, as already stated, Eq.~\eqref{eq:GMF} really 
corresponds to $\mathbb{G}_0(k,\omega)$ shifted by $J_0S$ in the case of $a$ 
band, and by $J_{0}S/2$ in the case of $b$ states. Therefore, we expect the 
MF solution~\eqref{eq:MF} to resemble the free hole dispersion, shifted 
to the lower energy range by the appropriate value, and with an energy 
gap of $J_{0}S/2$ --- indeed this is the 
case, as we show in a broad range of parameters in Sec.~\ref{sec:resu}.
We analyze there whether this prediction of the MF approximation holds 
beyond the regime of weak coupling $J_0\ll t$.

On the one hand, the two approximations described above are expected to
coincide with the exact solution in their respective parameter ranges.
Being among the most established approximate methods for quantum
many-body systems, they serve as benchmarks of the method used here.
On the other hand, comparing their predictions with the exact solution
in the intermediate parameter range, i.e. $J_0\sim t$ and $J_0\sim J$,
can give us a better understanding of how biased exactly those methods 
are. This is especially the case for the MF approach which
is often employed as a first attempt at tackling a complicated problem.

\section{Numerical results}
\label{sec:resu}

\begin{figure}[b!]
  \centering
  \includegraphics[width=7.7cm]{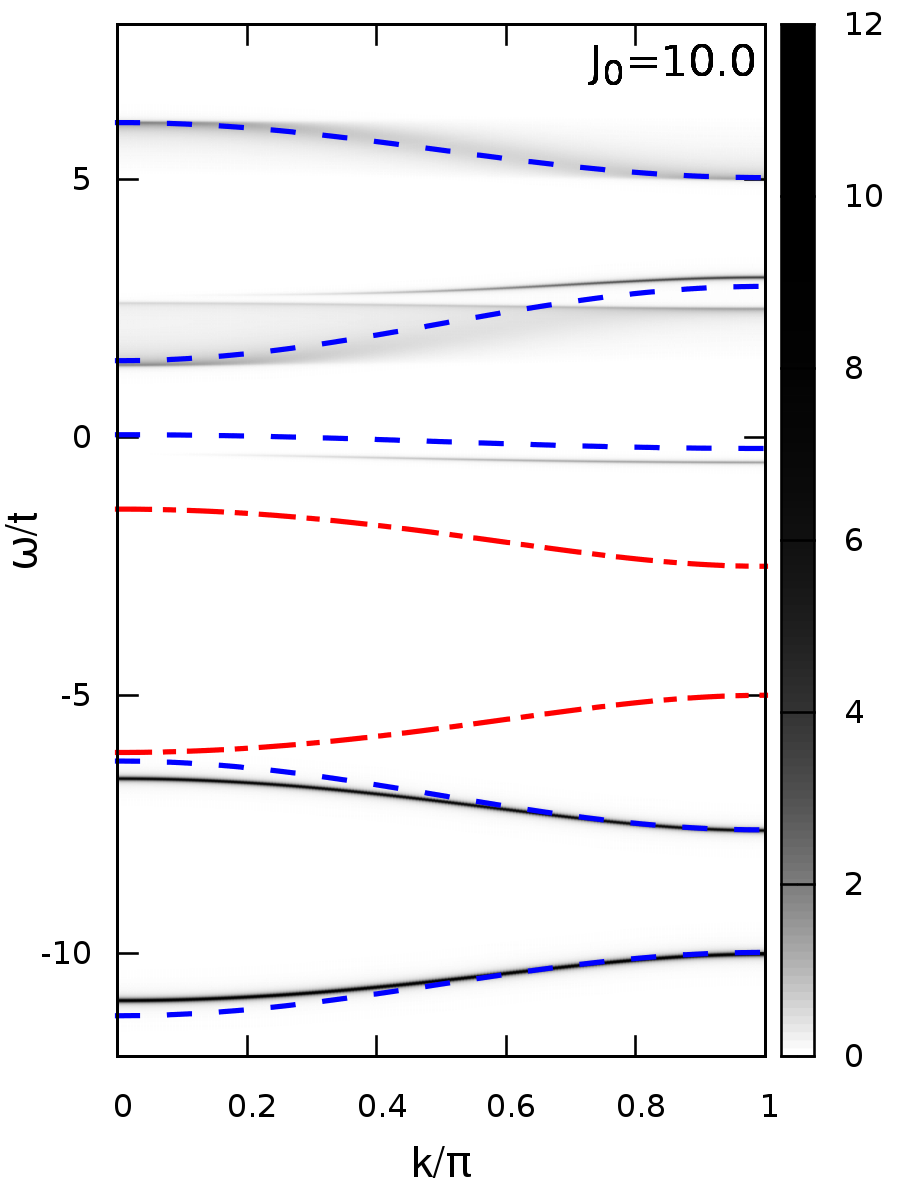}
  \caption{Spectral function $A(k,\omega)$ density map for $s$ orbital 
symmetry (shaded areas) compared with the analytic solutions 
\eqref{eq:eigenergy} obtained in perturbation theory in the strong 
coupling regime, shown by dashed (blue) lines. The dash-dotted (red) 
lines represent the MF states \eqref{eq:MF}. 
Parameters: $J_{0}=10t$, $J=0.05t$, $\eta=0.02t$, $S=1/2$.}
  \label{fig:perturb}
\end{figure}

\begin{figure*}[t!]
  \centering
  \includegraphics[width=0.9\textwidth]{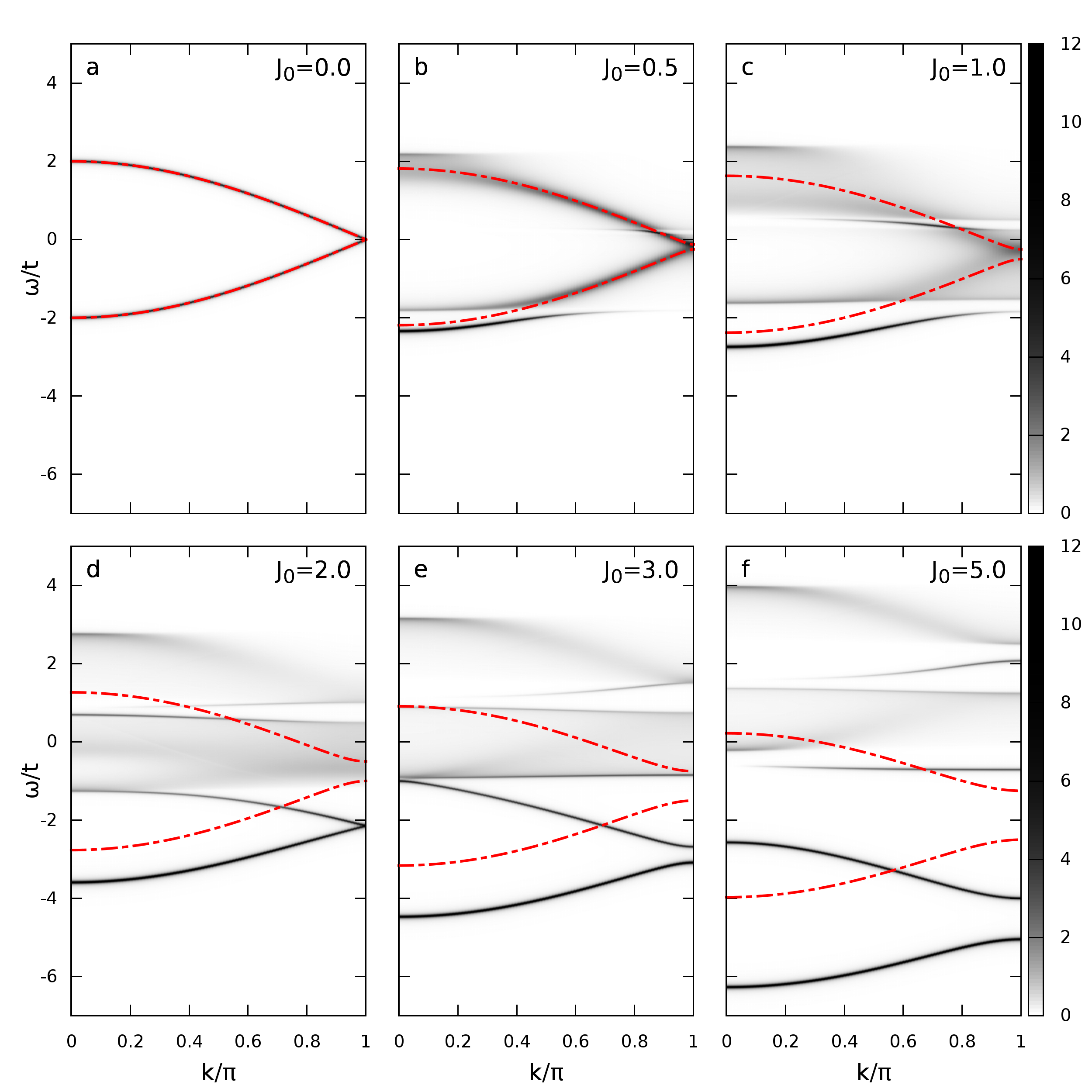}
\caption{Spectral function $A(k,\omega)$ density maps obtained for a 
hole with $\downarrow$-spin added to a FM chain (1) for $s$ orbital
symmetry. The dash-dotted (red) lines represent the MF states 
\eqref{eq:MF}. Note the strongly nonlinear scale of the map, employed 
in order to bring out the low-amplitude incoherent spectra.
Parameters: $J=0.05t$, $\eta=0.02t$, $S=1/2$.}
  \label{fig:spects}
\end{figure*}

The obtained result for the Green's function Eq.~\eqref{eq:G2} is exact, 
i.e., it follows from a rigorous derivation with no approximations 
employed, but unfortunately it does not allow one to calculate 
$\mathbb{G}(k,\omega)$ analytically.
In particular, its central part, the matrix $\mathbb{M}(k,\omega)$, 
has to be obtained numerically. Below we present the numerical results
obtained for the spectral function $A(k,\omega)$ using this exact
scheme. In the numerical calculations we take $t=1$ as the energy unit 
and set $J=0.05t$, $\eta=0.02t$. We consider the case of $S=1/2$, 
where quantum spin fluctuations are the most important. We then explore 
the dependence of the spectra on the value of the coupling constant 
$J_{0}$ which controls the strength of the interaction between localized 
spins and a hole in $\mathcal{V}$.

Let us consider first the strong coupling limit of $J_0=10t$, see
Fig.~\ref{fig:perturb}. In this regime one expects that the spectral 
function $A(k,\omega)$ consists of five features which correspond to
the perturbative states~\eqref{eq:eigstate}, with distinct energies 
and rather weak dispersion. This analytic result is confirmed by a
numerical solution, with the largest intensities obtained for the two 
states with the lowest energies. We remark that the states obtained in 
the perturbative regime have the same splitting of $J_0S/2$ at
$k=\pi$ as in the MF theory, but they appear at a much lower energy due 
to formation of polaron states. This demonstrates the importance 
of quantum spin fluctuations in the binding energies of these 
polaronic states, which are neglected in the MF approximation. Quantum 
spin fluctuations enhance the binding energy roughly by $J_0/2$. 

Consider next the systematic changes of the spectral functions with 
increasing exchange coupling $J_0$.
Fig.~\ref{fig:spects} shows the spectral function density maps for
the $s$ orbital symmetry for a wide range of $J_{0}$ values. For
intermediate values of $J_0$ it consists of distinct QP states and
shadded areas of scattering states. A nonlinear map scale has been 
applied in order to amplify the low-amplitude incoherent part of the 
spectrum, which in reality is negligibly small. The diversification of 
QP states caused by the interaction $\mathcal{V}$ can clearly be seen. 

Starting from $J_{0}=0$ two branches are seen, corresponding to the free 
hole propagation of a $\downarrow$-spin hole and exactly replicate the 
MF solution. Since in this situation there is no interaction whatsoever, 
the added charge (hole) propagates without coupling to the magnetic 
background.

Next, for $J_{0}=0.5t$ the two branches are seen to have widened
considerably and two new distinct features can be identified:
(i) one directly below the lower band and corresponding to the first 
polaronic state $\ket{\bar{a}}$, as shown by solutions obtained within 
perturbation theory and compared to the exact solution for large 
$J_{0}=10t$ (see Fig.~\ref{fig:perturb}), and 
(ii) the other one located slightly above $\omega=0$, and extending 
into the whole Brillouin zone for higher values of $J_{0}$.
This latter feature fades away considerably and gradually develops into 
the upper bound of the lower incoherent region, corresponding to 
$\ket{b}$ in the high coupling regime. At $J_0=t$ the original two branches 
have all but disappeared, and the lowest polaron state has almost fully 
developed. 

\begin{figure}[t!]
  \centering
  \includegraphics[width=\columnwidth]{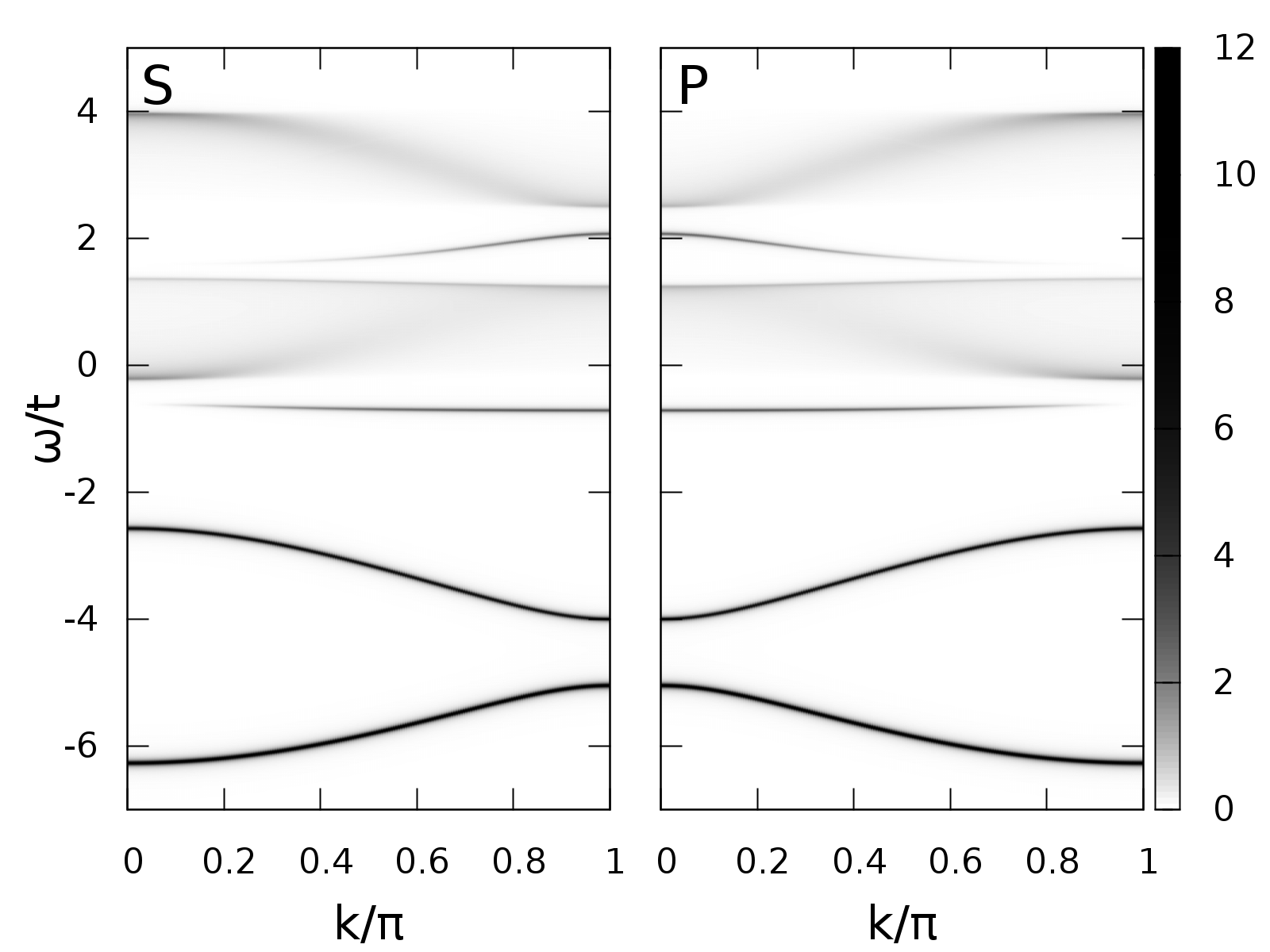}
\caption{Comparison of the exact solutions for the spectral function 
$A(k,\omega)$ Eq. \eqref{eq:trace} obtained for $s$ symmetry orbitals in
Fig. 1 (left) and for oxygen orbitals of $p$ symmetry (right).
Parameters: $J_{0}=5t$, $J=0.05t$, $\eta=0.02t$ and $S=1/2$.}
  \label{fig:SvsP}
\end{figure}

Increasing the interaction further to $J_{0}=2t$ we see that another 
state begins to emerge just slightly above the lowest polaronic state, 
starting from $k=\pi$. This state, corresponding to $\ket{\bar{b}}$ in 
the strong coupling regime then slowly develops while lowering further 
below the incoherent continuum from which it emerged. Around this point 
the incoherent part of the spectrum develops a gap and divides into two 
distinct parts --- the first of which has already been mentioned. 
The other one, situated at a higher energy, develops later into the 
$\ket{a}$ state. Finally, at $J_{0}=5t$ yet another state can be seen 
situated close to $\omega=0$, seemingly with no dispersion. This state 
can be identified as $\ket{m}$ and is purely magnonic, while the hole 
has the reversed $\uparrow$-spin.

While it is clear that MF gives good approximations for the weak
coupling regime, a curious observation can be made about the strong
coupling. Looking at the MF solutions plotted against the exact
results for large values of $J_{0}$, one notices a surprising 
resemblance to the two lowest-lying states $\ket{\bar{a}}$ and 
$\ket{\bar{b}}$, save for some constant energy shift. This indicates 
that MF approximation can give relatively good qualitative results, 
predicting correct dispersion for polaronic states, but introduces a
systematic error, as it neglects the binding energy coming from
hole-magnon interaction. This explains the huge discrepancy between MF 
energies and the exact energies found for the polaron states.

It is also interesting to note that, while MF predicts the gap to
develop monotonically, the real solution develops a gap shortly after
the $\ket{\bar{a}}$ state emerges from the incoherent region of the 
spectrum. This gap then closes again at around $J_0=2t$ and only after 
that does it reappear and start to widen monotonically. For more 
details on the evolution of the QP spectra with increasing parameter 
$J_0/t$, please refer to the Supplemental Material.\cite{suppl} 

\begin{figure}[t!]
  \centering
  \includegraphics[width=\columnwidth]{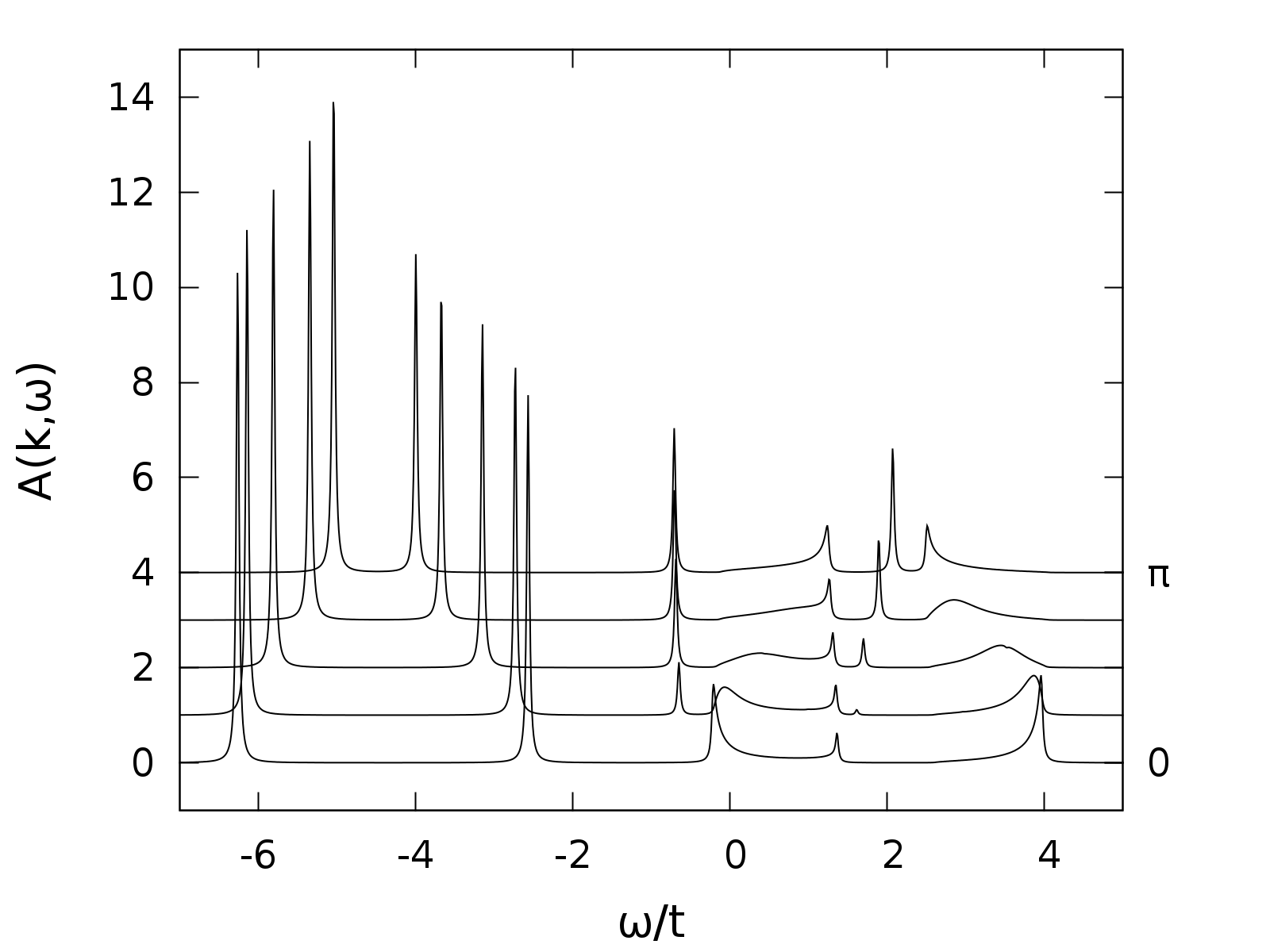}
\caption{The spectral function $A(k,\omega)$ for selected values of 
$k=0,0.25\pi,0.5\pi,0.25\pi,\pi$, shown also in Fig.~\ref{fig:spects}(f).
Parameters: $J_{0}=5t$, $J=0.05t$, $\eta=0.02t$ and $S=1/2$.}
  \label{fig:cuts}
\end{figure}

Apart from calculations made for a wide range of $J_{0}$ values for
$s$ symmetry, we have also done calculations for $p$ orbitals. However,
because $\mathcal{V}$ in our model does not distinguish between the
two, the only difference will come from their difference in dispersion.
Taking into account that $\sin(k/2)=\cos[(\pi-k)/2]$, we 
expect the solution for $p$ orbitals to be a ``mirror image'' of the
$s$-symmetry solutions with respect to momentum $k=\pi/2$.
Fig.~\ref{fig:SvsP} clearly demonstrates that this indeed is the case.

While the spectral maps are very useful in presenting the entire spectra 
obtained for the present excat solution, they do not give one a good 
sense of detail. For this reason, in Fig.~\ref{fig:cuts} we present an 
example of the spectra obtained for $J_{0}=5t$ for a few selected points 
$k\in[0,\pi]$ of the Brillouin zone. All the five spectral features
corresponding to the states $\ket{\bar{a}}_{k}$, $\ket{\bar{b}}_{k}$, 
$\ket{m}_k$, $\ket{b}_k$ and $\ket{a}_k$ can be well distinguished
from one another. The two lowest states clearly have the largest spectral 
weights.

\section{Discussion and summary}
\label{sec:summa}

We have used the method developed by M\"oller, Sawatzky and Berciu 
\cite{Mir12a,*Mir12b} to calculate the exact Green's function and the 
spectral function for a simple model of a single hole moving in a 
CuO$_{3}$-like FM chain. Five distinct spectral features are identified 
--- three of which arise from the hole propagating over the $a$ orbitals 
along the chain, and the other two follow from the hole within apical 
$b$ orbitals in a CuO$_3$-like chain. By introducing a realistic 
orbital structure for multi-band model, we addressed the problem of hole 
dynamics within $p$ orbitals in the charge-transfer model for a CuO$_3$
chain. We have then benchmarked this solution against the perturbation
theory at strong coupling and mean field approximations. We have found 
that both of these approaches coincide quite well with the Green's 
function solution in their respective regimes of applicability, i.e., 
mean field gives realistic predictions for weak interactions, while the 
perturbation theory reproduces all the states reasonably well in the 
strong coupling regime. The quantum states which develop beyond the 
mean field approximation will decrese their spectral weight with 
increasing value of spin $S$. In addition, the mean field approach seems 
to recreate the shapes of the polaronic bands at the strong coupling, 
but highly underrates the binding energy.

The perturbation solution allows us for identification of five distinct
states: two well defined binding polaronic bands and one nearly
dispersionless purely magnonic band, accompanied by two distinct 
excited polaronic states, coupled by a broad continuum. These latter 
excited states are much broader and have smaller spectral weights, 
even for very strong coupling $J_0\gg t$, which can be understood as 
following from the continuum of magnon excitations. Furthermore, the 
$\ket{b}_k,\ket{a}_k$ and $\ket{m}_k$ states which develop beyond 
the mean field approximation will decrease their spectral weight with 
increasing value of spin $S$ in the ferromagnetic chain. Indeed, the 
modifications of the spectra arising from quantum spin fluctuations are 
largest for $S=1/2$ and decrease with increasing $S$.

We also note that there is no essential difference between $s$ and $p$
orbital symmetries for the injected hole in the model Eq. 
\eqref{eq:model}, which is a result of taking into account only the 
exchange terms (second order two-site $p$-$d$-$p$ hopping). Therefore, 
such a simple model cannot properly describe a system with O-based 
conductance, reminiscent of doped cuprates. The simplest generalization 
of the present model is to include the three-site $p$-$d$-$p$ terms,
\cite{Zaa88} which distinguish orbital symmetry. This is an interesting
problem for future studies.

\acknowledgments

We thank Mona Berciu for insightful discussions. 
We kindly acknowledge financial support by the Polish National 
Science Center (NCN) under Project No. 2012/04/A/ST3/00331.

\appendix*

\section{Details of the exact solution Eq.~\eqref{eq:G2}}
\label{sec:appen}

Here we present the details of the calculation of the anomalous Green's
function $\tilde{\mathbb{F}}$. After applying the Dyson's
equation~\eqref{eq:dyson} to the Green's function appearing in the
definition given in Eq.~\eqref{eq:anomalous}, one obtains
\begin{equation}
\label{eq:anomalous2}
\begin{split}
  \mathbb{\tilde{F}}_{\mu\nu}(k,q,\omega) &= \MF
  \mu_{k\downarrow}\mathcal{G}(\omega)\mathcal{V}\nu_{k-q,\uparrow}^{\dag}
  S_{q}^{-} \FM\\
  &\times \mathbb{G}_{0}(k-q,\omega-\Omega_{q}),
\end{split}
\end{equation}
where the free-standing $\mathcal{G}_{0}(\omega)$ disappears due to the
anomalous average.

Since the total spin of the system is conserved, only one spin-flip is 
allowed in the FM background interacting with the hole. Therefore, in 
the present case $\mathcal{V}$ can only leave the same defected state 
(which causes a renormalization of $\tilde{\mathbb{F}}(k,q,\omega)$) or 
may reproduce the initial FM state by means of a deexcitation of a 
magnon by a term $\propto S^{+}s^{-}$. 
This leads directly to the following equation:
\begin{widetext}
\begin{align}
  \label{eq:F1}
  \tilde{\mathbb{F}}(k,q,\omega) &= \left[
    -\frac{J_{0}}{N}\sum_{p}\tilde{\mathbb{F}}(k,p,\omega) \mathbb{V}(q-p)
    +\frac{2J_{0}S}{N}\mathbb{G}(k,\omega)\mathbb{V}(q)
  \right]
  \mathbb{G}_{0}(k-q,\omega-\Omega_{q})\mathbb{Q}_{F}(k,q,\omega),\\
  \mathbb{Q}_{F}(k,q,\omega) &=\big[
  \mathbb{I}-J_{0}S\mathbb{V}_{0}\mathbb{G}_{0}(k-q,\omega-\Omega_{q})\big]^{-1}.
\end{align}
One has as well,
\begin{equation}
  \label{eq:trig}
  \mathbb{V}(q-p) =
  \begin{pmatrix}
    \cos\frac{q-p}{2} & 0\\
    0 & \frac{1}{2}
  \end{pmatrix} =
  \begin{pmatrix}
    \cos \frac{q}{2} & 0\\
    0 & \frac{1}{2}
  \end{pmatrix}
  \begin{pmatrix}
    \cos \frac{p}{2} & 0\\
    0 & \frac{1}{2}
  \end{pmatrix} +
  \begin{pmatrix}
    \sin \frac{q}{2} & 0\\
    0 & \frac{1}{2}
  \end{pmatrix}
  \begin{pmatrix}
    \sin \frac{p}{2} & 0\\
    0 & \frac{1}{2}
  \end{pmatrix} =
  \mathbb{V}(q)\mathbb{V}(p)+\bar{\mathbb{V}}(q)\bar{\mathbb{V}}(p),
\end{equation}
and one finds,
\begin{equation}
  \label{eq:F2}
  \tilde{\mathbb{F}}(k,q,\omega) = \left[
    -\frac{J_{0}}{N} (\mathbb{F}(k,\omega)\mathbb{V}(q)
    +\bar{\mathbb{F}}(k,\omega)\bar{\mathbb{V}}(q))
    +\frac{2J_{0}S}{N}\mathbb{G}(k,\omega)\mathbb{V}(q)
  \right]
  \mathbb{G}_{0}(k-q,\omega-\Omega_{q})\mathbb{Q}_{F}(k,q,\omega),
\end{equation}
\end{widetext}
where
\begin{equation}
  \label{eq:Fs}
  \bar{\mathbb{F}}(k,\omega) = \sum_{p}
  \tilde{\mathbb{F}}(k,p,\omega) \bar{\mathbb{V}}(p).
\end{equation}
$\mathbb{F}(k,\omega)$ used here is defined in Eq.~\eqref{eq:fmatrix}.
Multiplying now Eq.~\eqref{eq:F2} either by $\mathbb{V}(q)$ or by 
$\bar{\mathbb{V}}(q)$ and summing over $q$, one can find explicit
equations for $\mathbb{F}(k,\omega)$ and $\bar{\mathbb{F}}(k,\omega)$
in terms of $\mathbb{G}(k,\omega)$. Since $\bar{\mathbb{F}}(k,\omega)$
serves only as an auxiliary function, we will only present
$\mathbb{F}(k,\omega)$ here:
\begin{equation}
  \label{eq:F3}
  \mathbb{F}(k,\omega) =
  2S\mathbb{G}(k,\omega)[\mathbb{I}-\mathbb{M}^{-1}(k,\omega)],
\end{equation}
where we have already introduced the matrix $\mathbb{M}(k,\omega)$,
defined as follows:
\begin{equation}
  \label{eq:M}
  \begin{split}
    \mathbb{M}(k,\omega) &= \mathbb{I} +J_{0}\,\mathbb{G}_{cc}(k,\omega)
    -J_{0}^{2}\,\mathbb{G}_{cs}(k,\omega)\\
    &\times [\mathbb{I}+J_{0}\mathbb{G}_{ss}(k,\omega)]^{-1}
    \mathbb{G}_{sc}(k,\omega),
  \end{split}
\end{equation}
which is closely related to the self-energy. The auxiliary matrices 
introduced in Eq.~\eqref{eq:M} are:
\begin{subequations}
  \label{eq:Gcc}
  \begin{align}
    \mathbb{G}_{cc} &= \frac{1}{N}\sum_{q} \mathbb{V}(q)
    \mathbb{G}_{0}(k-q,\omega-\Omega_{q}) \mathbb{V}(q), \\
    \mathbb{G}_{cs} &= \frac{1}{N}\sum_{q} \mathbb{V}(q)
    \mathbb{G}_{0}(k-q,\omega-\Omega_{q}) \bar{\mathbb{V}}(q),\\
    \mathbb{G}_{sc} &= \frac{1}{N}\sum_{q} \bar{\mathbb{V}}(q)
    \mathbb{G}_{0}(k-q,\omega-\Omega_{q}) \mathbb{V}(q), \\
    \mathbb{G}_{ss} &= \frac{1}{N}\sum_{q} \bar{\mathbb{V}}(q)
    \mathbb{G}_{0}(k-q,\omega-\Omega_{q}) \bar{\mathbb{V}}(q).
  \end{align}
\end{subequations}
After plugging the solution Eq.~\eqref{eq:F3} into Eq.~\eqref{eq:G1}
and solving for $\mathbb{G}(k,\omega)$, one obtains the final result,
Eq.~\eqref{eq:G2} of Sec.~\ref{sec:green}.


\begin{thebibliography}{44}%
\makeatletter
\providecommand \@ifxundefined [1]{%
 \@ifx{#1\undefined}
}%
\providecommand \@ifnum [1]{%
 \ifnum #1\expandafter \@firstoftwo
 \else \expandafter \@secondoftwo
 \fi
}%
\providecommand \@ifx [1]{%
 \ifx #1\expandafter \@firstoftwo
 \else \expandafter \@secondoftwo
 \fi
}%
\providecommand \natexlab [1]{#1}%
\providecommand \enquote  [1]{``#1''}%
\providecommand \bibnamefont  [1]{#1}%
\providecommand \bibfnamefont [1]{#1}%
\providecommand \citenamefont [1]{#1}%
\providecommand \href@noop [0]{\@secondoftwo}%
\providecommand \href [0]{\begingroup \@sanitize@url \@href}%
\providecommand \@href[1]{\@@startlink{#1}\@@href}%
\providecommand \@@href[1]{\endgroup#1\@@endlink}%
\providecommand \@sanitize@url [0]{\catcode `\\12\catcode `\$12\catcode
  `\&12\catcode `\#12\catcode `\^12\catcode `\_12\catcode `\%12\relax}%
\providecommand \@@startlink[1]{}%
\providecommand \@@endlink[0]{}%
\providecommand \url  [0]{\begingroup\@sanitize@url \@url }%
\providecommand \@url [1]{\endgroup\@href {#1}{\urlprefix }}%
\providecommand \urlprefix  [0]{URL }%
\providecommand \Eprint [0]{\href }%
\providecommand \doibase [0]{http://dx.doi.org/}%
\providecommand \selectlanguage [0]{\@gobble}%
\providecommand \bibinfo  [0]{\@secondoftwo}%
\providecommand \bibfield  [0]{\@secondoftwo}%
\providecommand \translation [1]{[#1]}%
\providecommand \BibitemOpen [0]{}%
\providecommand \bibitemStop [0]{}%
\providecommand \bibitemNoStop [0]{.\EOS\space}%
\providecommand \EOS [0]{\spacefactor3000\relax}%
\providecommand \BibitemShut  [1]{\csname bibitem#1\endcsname}%
\let\auto@bib@innerbib\@empty
%</preamble>
\bibitem [{\citenamefont {Imada}\ \emph {et~al.}(1998)\citenamefont {Imada},
  \citenamefont {Fujimori},\ and\ \citenamefont {Tokura}}]{Ima98}%
  \BibitemOpen
  \bibfield  {author} {\bibinfo {author} {\bibfnamefont {M.}~\bibnamefont
  {Imada}}, \bibinfo {author} {\bibfnamefont {A.}~\bibnamefont {Fujimori}}, \
  and\ \bibinfo {author} {\bibfnamefont {Y.}~\bibnamefont {Tokura}},\ }\href
  {\doibase 10.1103/RevModPhys.70.1039} {\bibfield  {journal} {\bibinfo
  {journal} {Rev. Mod. Phys.}\ }\textbf {\bibinfo {volume} {70}},\ \bibinfo
  {pages} {1039} (\bibinfo {year} {1998})}\BibitemShut {NoStop}%
\bibitem [{\citenamefont {Zaanen}\ and\ \citenamefont {Ole\'s}(1993)}]{Zaa93}%
  \BibitemOpen
  \bibfield  {author} {\bibinfo {author} {\bibfnamefont {J.}~\bibnamefont
  {Zaanen}}\ and\ \bibinfo {author} {\bibfnamefont {A.~M.}\ \bibnamefont
  {Ole\'s}},\ }\href {\doibase 10.1103/PhysRevB.48.7197} {\bibfield  {journal}
  {\bibinfo  {journal} {Phys. Rev. B}\ }\textbf {\bibinfo {volume} {48}},\
  \bibinfo {pages} {7197} (\bibinfo {year} {1993})}\BibitemShut {NoStop}%
\bibitem [{\citenamefont {Dagotto}(1994)}]{Dag94}%
  \BibitemOpen
  \bibfield  {author} {\bibinfo {author} {\bibfnamefont {E.}~\bibnamefont
  {Dagotto}},\ }\href {\doibase 10.1103/RevModPhys.66.763} {\bibfield
  {journal} {\bibinfo  {journal} {Rev. Mod. Phys.}\ }\textbf {\bibinfo {volume}
  {66}},\ \bibinfo {pages} {763} (\bibinfo {year} {1994})}\BibitemShut
  {NoStop}%
\bibitem [{\citenamefont {Lee}\ \emph {et~al.}(2006)\citenamefont {Lee},
  \citenamefont {Nagaosa},\ and\ \citenamefont {Wen}}]{Nag06}%
  \BibitemOpen
  \bibfield  {author} {\bibinfo {author} {\bibfnamefont {P.~A.}\ \bibnamefont
  {Lee}}, \bibinfo {author} {\bibfnamefont {N.}~\bibnamefont {Nagaosa}}, \ and\
  \bibinfo {author} {\bibfnamefont {X.~G.}\ \bibnamefont {Wen}},\ }\href
  {\doibase 10.1103/RevModPhys.78.17} {\bibfield  {journal} {\bibinfo
  {journal} {Rev. Mod. Phys.}\ }\textbf {\bibinfo {volume} {78}},\ \bibinfo
  {pages} {17} (\bibinfo {year} {2006})}\BibitemShut {NoStop}%
\bibitem [{\citenamefont {Ogata}\ and\ \citenamefont {Fukuyama}(2008)}]{Oga08}%
  \BibitemOpen
  \bibfield  {author} {\bibinfo {author} {\bibfnamefont {M.}~\bibnamefont
  {Ogata}}\ and\ \bibinfo {author} {\bibfnamefont {H.}~\bibnamefont
  {Fukuyama}},\ }\href {\doibase 10.1088/0034-4885/71/3/036501} {\bibfield
  {journal} {\bibinfo  {journal} {Rep. Prog. Phys.}\ }\textbf {\bibinfo
  {volume} {71}},\ \bibinfo {pages} {1} (\bibinfo {year} {2008})}\BibitemShut
  {NoStop}%
\bibitem [{\citenamefont {Dagotto}\ \emph {et~al.}(2001)\citenamefont
  {Dagotto}, \citenamefont {Hotta},\ and\ \citenamefont {Moreo}}]{Dag01}%
  \BibitemOpen
  \bibfield  {author} {\bibinfo {author} {\bibfnamefont {E.}~\bibnamefont
  {Dagotto}}, \bibinfo {author} {\bibfnamefont {T.}~\bibnamefont {Hotta}}, \
  and\ \bibinfo {author} {\bibfnamefont {A.}~\bibnamefont {Moreo}},\ }\href
  {\doibase 10.1016/S0370-1573(00)00121-6} {\bibfield  {journal} {\bibinfo
  {journal} {Phys. Rep.}\ }\textbf {\bibinfo {volume} {344}},\ \bibinfo {pages}
  {1} (\bibinfo {year} {2001})}\BibitemShut {NoStop}%
\bibitem [{\citenamefont {Dagotto}(2005)}]{Dag05}%
  \BibitemOpen
  \bibfield  {author} {\bibinfo {author} {\bibfnamefont {E.}~\bibnamefont
  {Dagotto}},\ }\href {\doibase 10.1088/1367-2630/7/1/067} {\bibfield
  {journal} {\bibinfo  {journal} {New J. Phys.}\ }\textbf {\bibinfo {volume}
  {7}},\ \bibinfo {pages} {67} (\bibinfo {year} {2005})}\BibitemShut {NoStop}%
\bibitem [{\citenamefont {Wei\ss{}e}\ and\ \citenamefont
  {Fehske}(2004)}]{Wei04}%
  \BibitemOpen
  \bibfield  {author} {\bibinfo {author} {\bibfnamefont {A.}~\bibnamefont
  {Wei\ss{}e}}\ and\ \bibinfo {author} {\bibfnamefont {H.}~\bibnamefont
  {Fehske}},\ }\href@noop {} {\bibfield  {journal} {\bibinfo  {journal} {New J.
  Phys.}\ }\textbf {\bibinfo {volume} {6}},\ \bibinfo {pages} {158} (\bibinfo
  {year} {2004})}\BibitemShut {NoStop}%
\bibitem [{\citenamefont {Tokura}(2006)}]{Tok06}%
  \BibitemOpen
  \bibfield  {author} {\bibinfo {author} {\bibfnamefont {Y.}~\bibnamefont
  {Tokura}},\ }\href {\doibase 10.1088/0034-4885/69/3/R06} {\bibfield
  {journal} {\bibinfo  {journal} {Rep. Prog. Phys.}\ }\textbf {\bibinfo
  {volume} {69}},\ \bibinfo {pages} {797} (\bibinfo {year} {2006})}\BibitemShut
  {NoStop}%
\bibitem [{\citenamefont {de~Gennes}(1960)}]{deG60}%
  \BibitemOpen
  \bibfield  {author} {\bibinfo {author} {\bibfnamefont {P.~G.}\ \bibnamefont
  {de~Gennes}},\ }\href {\doibase 10.1103/PhysRev.118.141} {\bibfield
  {journal} {\bibinfo  {journal} {Phys. Rev.}\ }\textbf {\bibinfo {volume}
  {118}},\ \bibinfo {pages} {141} (\bibinfo {year} {1960})}\BibitemShut
  {NoStop}%
\bibitem [{\citenamefont {van~den Brink}\ and\ \citenamefont
  {Khomskii}(1999)}]{vdB99}%
  \BibitemOpen
  \bibfield  {author} {\bibinfo {author} {\bibfnamefont {J.}~\bibnamefont
  {van~den Brink}}\ and\ \bibinfo {author} {\bibfnamefont {D.}~\bibnamefont
  {Khomskii}},\ }\href@noop {} {\bibfield  {journal} {\bibinfo  {journal}
  {Phys. Rev. Lett.}\ }\textbf {\bibinfo {volume} {82}},\ \bibinfo {pages}
  {1016} (\bibinfo {year} {1999})}\BibitemShut {NoStop}%
\bibitem [{\citenamefont {Takenaka}\ \emph {et~al.}(2002)\citenamefont
  {Takenaka}, \citenamefont {Shiozaki},\ and\ \citenamefont {Sugai}}]{Tak02}%
  \BibitemOpen
  \bibfield  {author} {\bibinfo {author} {\bibfnamefont {K.}~\bibnamefont
  {Takenaka}}, \bibinfo {author} {\bibfnamefont {R.}~\bibnamefont {Shiozaki}},
  \ and\ \bibinfo {author} {\bibfnamefont {S.}~\bibnamefont {Sugai}},\
  }\href@noop {} {\bibfield  {journal} {\bibinfo  {journal} {Phys. Rev. B}\
  }\textbf {\bibinfo {volume} {65}},\ \bibinfo {pages} {184436} (\bibinfo
  {year} {2002})}\BibitemShut {NoStop}%
\bibitem [{\citenamefont {Ole\'s}\ and\ \citenamefont {Feiner}(2002)}]{Ole02}%
  \BibitemOpen
  \bibfield  {author} {\bibinfo {author} {\bibfnamefont {A.~M.}\ \bibnamefont
  {Ole\'s}}\ and\ \bibinfo {author} {\bibfnamefont {L.~F.}\ \bibnamefont
  {Feiner}},\ }\href {\doibase 10.1103/PhysRevB.65.052414} {\bibfield
  {journal} {\bibinfo  {journal} {Phys. Rev.~B}\ }\textbf {\bibinfo {volume}
  {65}},\ \bibinfo {pages} {052414} (\bibinfo {year} {2002})}\BibitemShut
  {NoStop}%
\bibitem [{\citenamefont {Feiner}\ and\ \citenamefont {Ole\'s}(2005)}]{Fei05}%
  \BibitemOpen
  \bibfield  {author} {\bibinfo {author} {\bibfnamefont {L.~F.}\ \bibnamefont
  {Feiner}}\ and\ \bibinfo {author} {\bibfnamefont {A.~M.}\ \bibnamefont
  {Ole\'s}},\ }\href {\doibase 10.1103/PhysRevB.71.144422} {\bibfield
  {journal} {\bibinfo  {journal} {Phys. Rev.~B}\ }\textbf {\bibinfo {volume}
  {71}},\ \bibinfo {pages} {144422} (\bibinfo {year} {2005})}\BibitemShut
  {NoStop}%
\bibitem [{\citenamefont {Ole\'s}(2012)}]{Ole12}%
  \BibitemOpen
  \bibfield  {author} {\bibinfo {author} {\bibfnamefont {A.~M.}\ \bibnamefont
  {Ole\'s}},\ }\href {\doibase 10.1088/0953-8984/24/31/313201} {\bibfield
  {journal} {\bibinfo  {journal} {J. Phys.: Condens. Matter}\ }\textbf
  {\bibinfo {volume} {24}},\ \bibinfo {pages} {313201} (\bibinfo {year}
  {2012})}\BibitemShut {NoStop}%
\bibitem [{\citenamefont {Emery}(1987)}]{Eme87}%
  \BibitemOpen
  \bibfield  {author} {\bibinfo {author} {\bibfnamefont {V.~J.}\ \bibnamefont
  {Emery}},\ }\href {\doibase 10.1103/PhysRevLett.58.2794} {\bibfield
  {journal} {\bibinfo  {journal} {Phys. Rev. Lett.}\ }\textbf {\bibinfo
  {volume} {58}},\ \bibinfo {pages} {2794} (\bibinfo {year}
  {1987})}\BibitemShut {NoStop}%
\bibitem [{\citenamefont {Varma}\ \emph {et~al.}(1987)\citenamefont {Varma},
  \citenamefont {S.~Schmitt-Rink},\ and\ \citenamefont {E.~Abrahams}}]{Var87}%
  \BibitemOpen
  \bibfield  {author} {\bibinfo {author} {\bibfnamefont {C.~M.}\ \bibnamefont
  {Varma}}, \bibinfo {author} {\bibfnamefont {S.}~\bibnamefont
  {S.~Schmitt-Rink}}, \ and\ \bibinfo {author} {\bibfnamefont {E.}~\bibnamefont
  {E.~Abrahams}},\ }\href@noop {} {\bibfield  {journal} {\bibinfo  {journal}
  {Solid State Commun.}\ }\textbf {\bibinfo {volume} {62}},\ \bibinfo {pages}
  {681} (\bibinfo {year} {1987})}\BibitemShut {NoStop}%
\bibitem [{\citenamefont {Ole\'s}\ \emph {et~al.}(1987)\citenamefont {Ole\'s},
  \citenamefont {Zaanen},\ and\ \citenamefont {Fulde}}]{Ole87}%
  \BibitemOpen
  \bibfield  {author} {\bibinfo {author} {\bibfnamefont {A.~M.}\ \bibnamefont
  {Ole\'s}}, \bibinfo {author} {\bibfnamefont {J.}~\bibnamefont {Zaanen}}, \
  and\ \bibinfo {author} {\bibfnamefont {P.}~\bibnamefont {Fulde}},\
  }\href@noop {} {\bibfield  {journal} {\bibinfo  {journal} {Physica B\&C}\
  }\textbf {\bibinfo {volume} {148}},\ \bibinfo {pages} {260} (\bibinfo {year}
  {1987})}\BibitemShut {NoStop}%
\bibitem [{\citenamefont {Zhang}\ and\ \citenamefont {Rice}(1988)}]{Zha88}%
  \BibitemOpen
  \bibfield  {author} {\bibinfo {author} {\bibfnamefont {F.~C.}\ \bibnamefont
  {Zhang}}\ and\ \bibinfo {author} {\bibfnamefont {T.~M.}\ \bibnamefont
  {Rice}},\ }\href@noop {} {\bibfield  {journal} {\bibinfo  {journal} {Phys.
  Rev.~B}\ }\textbf {\bibinfo {volume} {37}},\ \bibinfo {pages} {3759}
  (\bibinfo {year} {1988})}\BibitemShut {NoStop}%
\bibitem [{\citenamefont {Jefferson}\ \emph {et~al.}(1992)\citenamefont
  {Jefferson}, \citenamefont {Eskes},\ and\ \citenamefont {Feiner}}]{Jef92}%
  \BibitemOpen
  \bibfield  {author} {\bibinfo {author} {\bibfnamefont {J.~H.}\ \bibnamefont
  {Jefferson}}, \bibinfo {author} {\bibfnamefont {H.}~\bibnamefont {Eskes}}, \
  and\ \bibinfo {author} {\bibfnamefont {L.~F.}\ \bibnamefont {Feiner}},\
  }\href@noop {} {\bibfield  {journal} {\bibinfo  {journal} {Phys. Rev.~B}\
  }\textbf {\bibinfo {volume} {45}},\ \bibinfo {pages} {7959} (\bibinfo {year}
  {1992})}\BibitemShut {NoStop}%
\bibitem [{\citenamefont {Feiner}\ \emph {et~al.}(1996)\citenamefont {Feiner},
  \citenamefont {Jefferson},\ and\ \citenamefont {Raimondi}}]{Fei96}%
  \BibitemOpen
  \bibfield  {author} {\bibinfo {author} {\bibfnamefont {L.~F.}\ \bibnamefont
  {Feiner}}, \bibinfo {author} {\bibfnamefont {J.~H.}\ \bibnamefont
  {Jefferson}}, \ and\ \bibinfo {author} {\bibfnamefont {R.}~\bibnamefont
  {Raimondi}},\ }\href@noop {} {\bibfield  {journal} {\bibinfo  {journal}
  {Phys. Rev.~B}\ }\textbf {\bibinfo {volume} {53}},\ \bibinfo {pages} {8751}
  (\bibinfo {year} {1996})}\BibitemShut {NoStop}%
\bibitem [{\citenamefont {Raimondi}\ \emph {et~al.}(1996)\citenamefont
  {Raimondi}, \citenamefont {Jefferson},\ and\ \citenamefont {Feiner}}]{Rai96}%
  \BibitemOpen
  \bibfield  {author} {\bibinfo {author} {\bibfnamefont {R.}~\bibnamefont
  {Raimondi}}, \bibinfo {author} {\bibfnamefont {J.~H.}\ \bibnamefont
  {Jefferson}}, \ and\ \bibinfo {author} {\bibfnamefont {L.~F.}\ \bibnamefont
  {Feiner}},\ }\href@noop {} {\bibfield  {journal} {\bibinfo  {journal} {Phys.
  Rev.~B}\ }\textbf {\bibinfo {volume} {53}},\ \bibinfo {pages} {8774}
  (\bibinfo {year} {1996})}\BibitemShut {NoStop}%
\bibitem [{\citenamefont {Fujioka}\ \emph {et~al.}(2008)\citenamefont
  {Fujioka}, \citenamefont {Miyasaka},\ and\ \citenamefont {Tokura}}]{Fuj08}%
  \BibitemOpen
  \bibfield  {author} {\bibinfo {author} {\bibfnamefont {J.}~\bibnamefont
  {Fujioka}}, \bibinfo {author} {\bibfnamefont {S.}~\bibnamefont {Miyasaka}}, \
  and\ \bibinfo {author} {\bibfnamefont {Y.}~\bibnamefont {Tokura}},\ }\href
  {\doibase 10.1103/PhysRevB.77.144402} {\bibfield  {journal} {\bibinfo
  {journal} {Phys. Rev. B}\ }\textbf {\bibinfo {volume} {77}},\ \bibinfo
  {pages} {144402} (\bibinfo {year} {2008})}\BibitemShut {NoStop}%
\bibitem [{\citenamefont {Horsch}\ and\ \citenamefont {Ole\'s}(2011)}]{Hor11}%
  \BibitemOpen
  \bibfield  {author} {\bibinfo {author} {\bibfnamefont {P.}~\bibnamefont
  {Horsch}}\ and\ \bibinfo {author} {\bibfnamefont {A.~M.}\ \bibnamefont
  {Ole\'s}},\ }\href {\doibase 10.1103/PhysRevB.84.064429} {\bibfield
  {journal} {\bibinfo  {journal} {Phys. Rev. B}\ }\textbf {\bibinfo {volume}
  {84}},\ \bibinfo {pages} {064429} (\bibinfo {year} {2011})}\BibitemShut
  {NoStop}%
\bibitem [{\citenamefont {Sakai}\ \emph {et~al.}(2010)\citenamefont {Sakai},
  \citenamefont {Ishiwata}, \citenamefont {Okuyama}, \citenamefont {Nakao},
  \citenamefont {Nakao}, \citenamefont {Murakami}, \citenamefont {Taguchi},\
  and\ \citenamefont {Tokura}}]{Sak10}%
  \BibitemOpen
  \bibfield  {author} {\bibinfo {author} {\bibfnamefont {H.}~\bibnamefont
  {Sakai}}, \bibinfo {author} {\bibfnamefont {S.}~\bibnamefont {Ishiwata}},
  \bibinfo {author} {\bibfnamefont {D.}~\bibnamefont {Okuyama}}, \bibinfo
  {author} {\bibfnamefont {A.}~\bibnamefont {Nakao}}, \bibinfo {author}
  {\bibfnamefont {H.}~\bibnamefont {Nakao}}, \bibinfo {author} {\bibfnamefont
  {Y.}~\bibnamefont {Murakami}}, \bibinfo {author} {\bibfnamefont
  {Y.}~\bibnamefont {Taguchi}}, \ and\ \bibinfo {author} {\bibfnamefont
  {Y.}~\bibnamefont {Tokura}},\ }\href {\doibase 10.1103/PhysRevB.82.180409}
  {\bibfield  {journal} {\bibinfo  {journal} {Phys. Rev. B}\ }\textbf {\bibinfo
  {volume} {82}},\ \bibinfo {pages} {180409} (\bibinfo {year}
  {2010})}\BibitemShut {NoStop}%
\bibitem [{\citenamefont {Ole\'s}\ and\ \citenamefont
  {Khaliullin}(2011)}]{Ole11}%
  \BibitemOpen
  \bibfield  {author} {\bibinfo {author} {\bibfnamefont {A.~M.}\ \bibnamefont
  {Ole\'s}}\ and\ \bibinfo {author} {\bibfnamefont {G.}~\bibnamefont
  {Khaliullin}},\ }\href {\doibase 10.1103/PhysRevB.84.214414} {\bibfield
  {journal} {\bibinfo  {journal} {Phys. Rev. B}\ }\textbf {\bibinfo {volume}
  {84}},\ \bibinfo {pages} {214414} (\bibinfo {year} {2011})}\BibitemShut
  {NoStop}%
\bibitem [{\citenamefont {Trugman}(1988)}]{Tru88}%
  \BibitemOpen
  \bibfield  {author} {\bibinfo {author} {\bibfnamefont {S.~A.}\ \bibnamefont
  {Trugman}},\ }\href {\doibase 10.1103/PhysRevB.37.1597} {\bibfield  {journal}
  {\bibinfo  {journal} {Phys. Rev. B}\ }\textbf {\bibinfo {volume} {37}},\
  \bibinfo {pages} {1597} (\bibinfo {year} {1988})}\BibitemShut {NoStop}%
\bibitem [{\citenamefont {Mart\'inez}\ and\ \citenamefont
  {Horsch}(1991)}]{Mar91}%
  \BibitemOpen
  \bibfield  {author} {\bibinfo {author} {\bibfnamefont {G.}~\bibnamefont
  {Mart\'inez}}\ and\ \bibinfo {author} {\bibfnamefont {P.}~\bibnamefont
  {Horsch}},\ }\href {\doibase 10.1103/PhysRevB.44.317} {\bibfield  {journal}
  {\bibinfo  {journal} {Phys. Rev.~B}\ }\textbf {\bibinfo {volume} {44}},\
  \bibinfo {pages} {317} (\bibinfo {year} {1991})}\BibitemShut {NoStop}%
\bibitem [{\citenamefont {Nolting}(1979)}]{Nol79}%
  \BibitemOpen
  \bibfield  {author} {\bibinfo {author} {\bibfnamefont {W.}~\bibnamefont
  {Nolting}},\ }\href {\doibase 10.1002/pssb.2220960102} {\bibfield  {journal}
  {\bibinfo  {journal} {Phys. Status Solidi B}\ }\textbf {\bibinfo {volume}
  {96}},\ \bibinfo {pages} {11} (\bibinfo {year} {1979})}\BibitemShut {NoStop}%
\bibitem [{\citenamefont {Nolting}\ and\ \citenamefont {Ole\'s}(1980)}]{Nol80}%
  \BibitemOpen
  \bibfield  {author} {\bibinfo {author} {\bibfnamefont {W.}~\bibnamefont
  {Nolting}}\ and\ \bibinfo {author} {\bibfnamefont {A.~M.}\ \bibnamefont
  {Ole\'s}},\ }\href {\doibase 10.1103/PhysRevB.22.6184} {\bibfield  {journal}
  {\bibinfo  {journal} {Phys. Rev.~B}\ }\textbf {\bibinfo {volume} {22}},\
  \bibinfo {pages} {6184} (\bibinfo {year} {1980})}\BibitemShut {NoStop}%
\bibitem [{\citenamefont {Nolting}\ and\ \citenamefont {Ole\'s}(1981)}]{Nol81}%
  \BibitemOpen
  \bibfield  {author} {\bibinfo {author} {\bibfnamefont {W.}~\bibnamefont
  {Nolting}}\ and\ \bibinfo {author} {\bibfnamefont {A.~M.}\ \bibnamefont
  {Ole\'s}},\ }\href {\doibase 10.1103/PhysRevB.23.4122} {\bibfield  {journal}
  {\bibinfo  {journal} {Phys. Rev.~B}\ }\textbf {\bibinfo {volume} {23}},\
  \bibinfo {pages} {4122} (\bibinfo {year} {1981})}\BibitemShut {NoStop}%
\bibitem [{\citenamefont {Wegner}\ \emph {et~al.}(1998)\citenamefont {Wegner},
  \citenamefont {Potthoff},\ and\ \citenamefont {Nolting}}]{Weg98}%
  \BibitemOpen
  \bibfield  {author} {\bibinfo {author} {\bibfnamefont {T.}~\bibnamefont
  {Wegner}}, \bibinfo {author} {\bibfnamefont {M.}~\bibnamefont {Potthoff}}, \
  and\ \bibinfo {author} {\bibfnamefont {W.}~\bibnamefont {Nolting}},\
  }\href@noop {} {\bibfield  {journal} {\bibinfo  {journal} {Phys. Rev.~B}\
  }\textbf {\bibinfo {volume} {57}},\ \bibinfo {pages} {6211} (\bibinfo {year}
  {1998})}\BibitemShut {NoStop}%
\bibitem [{\citenamefont {Nolting}\ \emph {et~al.}(1996)\citenamefont
  {Nolting}, \citenamefont {Jaya},\ and\ \citenamefont {Rex}}]{Nol96}%
  \BibitemOpen
  \bibfield  {author} {\bibinfo {author} {\bibfnamefont {W.}~\bibnamefont
  {Nolting}}, \bibinfo {author} {\bibfnamefont {S.~M.}\ \bibnamefont {Jaya}}, \
  and\ \bibinfo {author} {\bibfnamefont {S.}~\bibnamefont {Rex}},\ }\href
  {\doibase 10.1103/PhysRevB.54.14455} {\bibfield  {journal} {\bibinfo
  {journal} {Phys. Rev.~B}\ }\textbf {\bibinfo {volume} {54}},\ \bibinfo
  {pages} {14455} (\bibinfo {year} {1996})}\BibitemShut {NoStop}%
\bibitem [{\citenamefont {Daghofer}\ \emph {et~al.}(2004)\citenamefont
  {Daghofer}, \citenamefont {Ole\'s},\ and\ \citenamefont {von~der
  Linden}}]{Dag04}%
  \BibitemOpen
  \bibfield  {author} {\bibinfo {author} {\bibfnamefont {M.}~\bibnamefont
  {Daghofer}}, \bibinfo {author} {\bibfnamefont {A.~M.}\ \bibnamefont
  {Ole\'s}}, \ and\ \bibinfo {author} {\bibfnamefont {W.}~\bibnamefont {von~der
  Linden}},\ }\href {\doibase 10.1103/PhysRevB.70.184430} {\bibfield  {journal}
  {\bibinfo  {journal} {Phys. Rev. B}\ }\textbf {\bibinfo {volume} {70}},\
  \bibinfo {pages} {184430} (\bibinfo {year} {2004})}\BibitemShut {NoStop}%
\bibitem [{\citenamefont {Daghofer}\ \emph {et~al.}(2006)\citenamefont
  {Daghofer}, \citenamefont {Horsch},\ and\ \citenamefont
  {Khaliullin}}]{Dag06}%
  \BibitemOpen
  \bibfield  {author} {\bibinfo {author} {\bibfnamefont {M.}~\bibnamefont
  {Daghofer}}, \bibinfo {author} {\bibfnamefont {P.}~\bibnamefont {Horsch}}, \
  and\ \bibinfo {author} {\bibfnamefont {G.}~\bibnamefont {Khaliullin}},\
  }\href {\doibase 10.1103/PhysRevLett.96.216404} {\bibfield  {journal}
  {\bibinfo  {journal} {Phys. Rev. Lett.}\ }\textbf {\bibinfo {volume} {96}},\
  \bibinfo {pages} {216404} (\bibinfo {year} {2006})}\BibitemShut {NoStop}%
\bibitem [{\citenamefont {Avella}\ \emph {et~al.}(2013)\citenamefont {Avella},
  \citenamefont {Horsch},\ and\ \citenamefont {Ole\'s}}]{Ave13}%
  \BibitemOpen
  \bibfield  {author} {\bibinfo {author} {\bibfnamefont {A.}~\bibnamefont
  {Avella}}, \bibinfo {author} {\bibfnamefont {P.}~\bibnamefont {Horsch}}, \
  and\ \bibinfo {author} {\bibfnamefont {A.~M.}\ \bibnamefont {Ole\'s}},\
  }\href@noop {} {\bibfield  {journal} {\bibinfo  {journal} {Phys. Rev.~B}\
  }\textbf {\bibinfo {volume} {87}},\ \bibinfo {pages} {045132} (\bibinfo
  {year} {2013})}\BibitemShut {NoStop}%
\bibitem [{\citenamefont {Zaanen}\ and\ \citenamefont {Ole\'s}(1988)}]{Zaa88}%
  \BibitemOpen
  \bibfield  {author} {\bibinfo {author} {\bibfnamefont {J.}~\bibnamefont
  {Zaanen}}\ and\ \bibinfo {author} {\bibfnamefont {A.~M.}\ \bibnamefont
  {Ole\'s}},\ }\href {\doibase 10.1103/PhysRevB.37.9423} {\bibfield  {journal}
  {\bibinfo  {journal} {Phys. Rev.~B}\ }\textbf {\bibinfo {volume} {37}},\
  \bibinfo {pages} {9423} (\bibinfo {year} {1988})}\BibitemShut {NoStop}%
\bibitem [{\citenamefont {Prelov\v{s}ek}(1988)}]{Pre88}%
  \BibitemOpen
  \bibfield  {author} {\bibinfo {author} {\bibfnamefont {P.}~\bibnamefont
  {Prelov\v{s}ek}},\ }\href {\doibase 10.1016/0375-9601(88)90764-5} {\bibfield
  {journal} {\bibinfo  {journal} {Phys. Lett. A}\ }\textbf {\bibinfo {volume}
  {126}},\ \bibinfo {pages} {287} (\bibinfo {year} {1988})}\BibitemShut
  {NoStop}%
\bibitem [{\citenamefont {M\"oller}\ \emph
  {et~al.}(2012{\natexlab{a}})\citenamefont {M\"oller}, \citenamefont
  {Sawatzky},\ and\ \citenamefont {Berciu}}]{Mir12a}%
  \BibitemOpen
  \bibfield  {author} {\bibinfo {author} {\bibfnamefont {M.}~\bibnamefont
  {M\"oller}}, \bibinfo {author} {\bibfnamefont {G.~A.}\ \bibnamefont
  {Sawatzky}}, \ and\ \bibinfo {author} {\bibfnamefont {M.}~\bibnamefont
  {Berciu}},\ }\href {\doibase 10.1103/PhysRevLett.108.216403} {\bibfield
  {journal} {\bibinfo  {journal} {Phys. Rev. Lett.}\ }\textbf {\bibinfo
  {volume} {108}},\ \bibinfo {pages} {216403} (\bibinfo {year}
  {2012}{\natexlab{a}})}\BibitemShut {NoStop}%
\bibitem [{\citenamefont {M\"oller}\ \emph
  {et~al.}(2012{\natexlab{b}})\citenamefont {M\"oller}, \citenamefont
  {Sawatzky},\ and\ \citenamefont {Berciu}}]{Mir12b}%
  \BibitemOpen
  \bibfield  {author} {\bibinfo {author} {\bibfnamefont {M.}~\bibnamefont
  {M\"oller}}, \bibinfo {author} {\bibfnamefont {G.~A.}\ \bibnamefont
  {Sawatzky}}, \ and\ \bibinfo {author} {\bibfnamefont {M.}~\bibnamefont
  {Berciu}},\ }\href {\doibase 10.1103/PhysRevB.86.075128} {\bibfield
  {journal} {\bibinfo  {journal} {Phys. Rev. B}\ }\textbf {\bibinfo {volume}
  {86}},\ \bibinfo {pages} {075128} (\bibinfo {year}
  {2012}{\natexlab{b}})}\BibitemShut {NoStop}%
\bibitem [{\citenamefont {Schlappa~et al.}(2011)}]{Sch11}%
  \BibitemOpen
  \bibfield  {author} {\bibinfo {author} {\bibfnamefont {J.}~\bibnamefont
  {Schlappa~et al.}},\ }\href@noop {} {\bibfield  {journal} {\bibinfo
  {journal} {Nature}\ }\textbf {\bibinfo {volume} {485}},\ \bibinfo {pages}
  {82} (\bibinfo {year} {2011})}\BibitemShut {NoStop}%
\bibitem [{\citenamefont {Wohlfeld}\ \emph {et~al.}(2011)\citenamefont
  {Wohlfeld}, \citenamefont {Daghofer}, \citenamefont {Nishimoto},
  \citenamefont {Khaliullin},\ and\ \citenamefont {van~den Brink}}]{Woh11}%
  \BibitemOpen
  \bibfield  {author} {\bibinfo {author} {\bibfnamefont {K.}~\bibnamefont
  {Wohlfeld}}, \bibinfo {author} {\bibfnamefont {M.}~\bibnamefont {Daghofer}},
  \bibinfo {author} {\bibfnamefont {S.}~\bibnamefont {Nishimoto}}, \bibinfo
  {author} {\bibfnamefont {G.}~\bibnamefont {Khaliullin}}, \ and\ \bibinfo
  {author} {\bibfnamefont {J.}~\bibnamefont {van~den Brink}},\ }\href {\doibase
  10.1103/PhysRevLett.107.147201} {\bibfield  {journal} {\bibinfo  {journal}
  {Phys. Rev. Lett.}\ }\textbf {\bibinfo {volume} {107}},\ \bibinfo {pages}
  {147201} (\bibinfo {year} {2011})}\BibitemShut {NoStop}%
\bibitem [{\citenamefont {Berciu}\ and\ \citenamefont
  {Sawatzky}(2009)}]{berciu09}%
  \BibitemOpen
  \bibfield  {author} {\bibinfo {author} {\bibfnamefont {M.}~\bibnamefont
  {Berciu}}\ and\ \bibinfo {author} {\bibfnamefont {G.~A.}\ \bibnamefont
  {Sawatzky}},\ }\href {\doibase 10.1103/PhysRevB.79.195116} {\bibfield
  {journal} {\bibinfo  {journal} {Phys. Rev.~B}\ }\textbf {\bibinfo {volume}
  {79}},\ \bibinfo {pages} {195116} (\bibinfo {year} {2009})}\BibitemShut
  {NoStop}%
\bibitem [{sup()}]{suppl}%
  \BibitemOpen
  \href@noop {} {}\bibinfo {note} {See Supplemental Material at \url{} for an
  animated GIF, visualising the evolution of the spectra with increasing value
  of the Kondo-like exchange $J_{0}$.}\BibitemShut {Stop}%
\end{thebibliography}
\end{document}